\documentclass{aa}  
\usepackage{graphicx}
\usepackage{natbib}
\usepackage{pdfpages}
\usepackage{color}
\usepackage[varg]{txfonts}
\usepackage{amsmath}

\def\kms {\rm{km~s^{-1}}}
\def\apj {ApJ}
\def\apjl {ApJL}
\def\apjs {ApJS}
\def\aj {AJ}
\def\aap {A\&A}
\def\mnras {MNRAS}

\def\nat {Nature}

\begin{document}

   \title{Effects of environment on stellar metallicity profiles of late-type galaxies
in the CALIFA survey}


   \author{Valeria Coenda\inst{1,2}
          \and Dami\'an Mast\inst{2,3}
          \and Hern\'an Muriel\inst{1,2}
          \and H\'ector J. Mart\'inez\inst{1,2}}

   \institute{Instituto de Astronom\'ia Te\'orica y Experimental (IATE), CONICET - UNC, Laprida 854,
   X5000BGR,C\'ordoba, Argentina\\
             \and
             Observatorio Astron\'omico, Universidad Nacional de C\'ordoba, Laprida 854, X5000BGR, C\'ordoba, 
             Argentina.\\
             \and 
             Consejo de Investigaciones Cient\'{i}ficas y T\'ecnicas de la Rep\'ublica Argentina, Avda. Rivadavia 1917, C1033AAJ, CABA, Argentina.\\
              }

   \date{Received XXXX; accepted XXXX}

 
  \abstract
   {}
   {We explore the effects of environment in the evolution of late-type galaxies by studying the radial profiles of light- and mass-weighted metallicities of galaxies in two discrete environments: field and groups.}
   {
   To study the dependence of metallicity on environment. 
   We use a sample of 167 late-type galaxies, with stellar masses $9\le \log(M_{\star}/M_{\odot}) \le 12$,  drawn from the CALIFA survey.
   Firstly, we obtain light- and mass-weighted stellar metallicty profiles, and stellar mass density profiles of these galaxies, using publicly available data. 
   Then we classify them according to their environment into field and group
   galaxies. Finally, we make a study of the metallicity of galaxies
   in these two environments which includes the comparison of the metallicity as a function of the radius, at a characteristic scale, and as a function of a the stellar mass surface density. Since metallicity depends on galaxy mass, we take special care throughout the paper in order to compare, in all cases, subsamples of galaxies in groups and in the field that have similar masses. 
   }
   {We find significant differences between group and field late-type galaxies in terms of their metallicity, in the sense that group galaxies are systematically more metallic than their field counterparts. We find that field galaxies have, in general, metallicity profiles that show a negative gradient in their inner regions, and a shallower profile at larger radii. This contrasts with the metallicity profiles of galaxies in groups, which tend to be flat in the 
   inner regions, and to have a negative gradient in the outer parts.
   Regarding the metallicity at the
   characteristic radius of the luminosity profiles, we consistently find that it is higher for group galaxies irrespective of galaxy mass.  At fixed local 
   stellar surface mass density, group galaxies are again more metallic, also the dependence of metallicity on surface density is less important for group galaxies.
   }
   {The evidence of a clear difference on the metallicity of group and field galaxies, as a function of mass, spatial scale, and local stellar mass density,
   are indicative of the different evolutionary paths that galaxies in groups and
   in the field have followed. We discuss possible implications of the observed
   differences.}


   \maketitle
%

\section{Introduction}\label{intro}

The formation and evolution of galaxies is a complicated process that involves 
the action of different physical mechanisms, acting at different temporal, and 
spatial scales. As for their origin, these processes can be due to internal 
or to external, i.e., environmental, factors.
There are many internal physical mechanisms that can affect the properties of
galaxies, for example, supernovae (SN) outflows (e.g. \citealt{Stringer:2012}, 
\citealt{Bower:2012}), feedback from massive stars (e.g. 
\citealt{DallaV:2008}, \citealt{Hopkins:2012}), active galactic nuclei (AGN)
feedback (e.g. 
\citealt{Nandra:2007}, \citealt{Hasinger:2008}, \citealt{Silverman:2008}, 
\citealt{Cimatti:2013}), halo heating \citep{Marasco:2012}, and morphological 
quenching \citep{Martig:2009}. On the other hand, several environmental mechanisms 
act upon galaxies at different stages of their life. Galaxies in groups and clusters 
can loose an important fraction of their cold gas due to the pressure of the 
intracluster hot gas, a process known as ram pressure stripping (e.g. 
\citealt{GG:1972}, \citealt{Abadi:1999}, \citealt{Rasmussen:2006}, 
\citealt{Jaffe:2012}, \citealt{Hess:2013}).
The hot gas can also be removed from the galactic
halo, cutting off the supply of gas and consequently stopping star formation, 
this mechanism is known as starvation (e.g. \citealt{Larson:1980}, 
\citealt{Bekki:2009}, \citealt{McCarthy:2008}, \citealt{Bahe:2013}, 
\citealt{Vijayaraghavan:2015}). Other mechanisms such as tidal stripping (e.g. 
\citealt{Gnedin:2003a}, \citealt{Villalobos:2014}), and thermal evaporation 
\citep{Cowie:1977}, could be also responsible of affecting galaxy evolution. 
Galaxy-galaxy high speed interactions (e.g. \citealt{Moore:1996}, 
\citealt{Moore:1999},  \citealt{Gnedin:2003b}), and mergers, are mechanisms that can
redistribute the gaseous, dark matter, and stellar components of a galaxy, 
with the resulting change in its properties.  

Historically, statistical studies of galaxies have been carried out by analysing 
their integrated properties, such as star formation, metallicity, colour, magnitudes, 
etc. In recent years, thanks to the new generation of Integral Field 
Spectroscopy (IFS) surveys such as CALIFA \citep{CALIFA}, SAMI \citep{SAMI}, and 
MANGA \citep{MANGA}, it has become possible to obtain a spatially resolved 
information of the stellar population in galaxies. These
instruments have enabled the construction of surveys that include hundreds of 
galaxies and provide two-dimensional maps for different properties of galaxies, 
being the metallicity one of the most studied. The analysis of the spatial
distribution of the metallicity inside galaxies, is an important tool to study
different physical processes that act at different radii. 

Studies of the metallicity distribution in galaxies have been carried out following 
different approaches and techniques, analysing the gaseous and/or the stellar 
component. Some of these works have addressed the dependence of metallicity profiles on the environment. However, regardless of the techniques and galaxy samples used, the results continue 
to be contradictory \citep{Pilyugin2016}, and range from finding no difference
within 0.02 dex between field and cluster galaxies \citep{Hughes2013,Kacprzak2015}, to finding that galaxies in clusters are more metallic than 
those in the field \citep{Shimakawa2015}, or even that star forming (SF) galaxies
in clusters are less metallic than those in the field \citep{Valentino2015}. 
In general, the preferred method to determine the metallicity of SF galaxies, is to
measure the oxygen abundance (O/H) in the interstellar medium (ISM) because it is the most abundant heavy element, easy to detect and to
obtain measurements for a large sample of galaxies through their emission lines. Without discarding the historical factor that, 
for the previous reason, make the calibration methods of the abundance relations to 
be developed and improved with more and better quality data through the years. 
However, the different calibrators available to determine the gas-phase metallicity \citep[e.g.][]{Perez-Montero2005, Marino2013,Perez-Montero2017} have a considerable scatter and it may not be easy to determine 
whether the variations due to the effects of the environment are greater than this 
scatter. Many studies \citep[e.g.][]{Ellison:2009}, prefer gas-phase metallicity 
because stellar metallicity is insensitive to small variations in metallicity. 
These authors find variations of 0.02-0.07 dex between different environments.
The stellar metallicity accounts for the chemical enrichment throughout 
the star-forming history of the galaxy. Although HII regions have a memory of the past SFH \citep{Sanchez2014}, gas-phase metallicity is susceptible to outflows and inflows, among others process \citep{Lian2017,Wu2017}, that can alter the metallicity radial profiles. For this reason, in this paper we have focused on stellar 
metallicity. Although the expected variations are supposed to be small, 
we are interested in the evolutionary imprint left by processes acting in different 
environments. These processes may act on shorter time scales at the ISM, but the 
imprint of their effect, as far as galactic evolution is concerned, will eventually 
be on the stellar metallicity.  
It is important to note that, due to the small variations expected from environmental 
effects, and given the different processes acting at different radii depending on the 
morphological type and mass of the galaxies, the best way to study these effects is 
through a spatially resolved analysis, i.e., a study of integrated properties, as 
opposed to IFS, may blur any existing evidence of the acting mechanisms.

Several works that use numerical simulations to study the formation and evolution
of disks in galaxies favour an inside-out scenario for this component
(e.g. \citealt{Scannapieco:2009,Brook:2012,Tissera:2016}).
Many observational results using IFS provide support for this.
Using the CALIFA survey, \citet{SanchezBlazquez:2014} find  shallow
and negative metallicity gradients in disk galaxies. 
In particular, they find that luminosity-weighted metallicity 
gradients are steeper than the mass-weighted ones. They also analyse whether the 
presence of a bar may originate the shallow profiles observed, since bars
are supposed to produce stellar migrations (e.g. \citealt{Wielen:1977}, 
\citealt{Sellwood:2002}). However, they do not find significant differences 
between barred and non-barred galaxies, in disagreement with the  predictions of
numerical simulations (e.g.  \citealt{Minchev:2012,DiMatteo:2013,Vincenzo:2020}). 
\citet{GonzalezDelgado:2016} also find that spiral galaxies have negative metallicity 
and age gradients, in agreement with an inside-out formation. 
Further evidence in favour of the inside-out scenario using CALIFA data is 
presented in \citet{Garcia-Benito:2017}. They study the mass assembly time scales 
of 661 CALIFA galaxies that cover wide ranges in both, mass, and Hubble types.
Their results indicate that galaxies form inside-out independently of their stellar 
mass, stellar mass surface density, and morphology.
\citet{Lian:2018}, using data from the MaNGA survey, also report negative gradients 
in both, gas, and stellar metallicity, the latter being steeper.

\citet{Goddard:2017b} use the MANGA survey to study the internal gradients of the
stellar population in galaxies. They obtain negative metallicity gradients for both, 
early, and late-type galaxies. They find that gradients are steeper for late-types.
In addition, \citet{Goddard:2017a} analyzed the stellar population properties, age and metallicity, to study the gradients as a 
function of three characterisations of the environment: local density, tidal strength 
parameter, and whether a galaxy is central or satellite. 
In neither case, they find a strong correlation with the environment, and 
suggest that galaxy mass is the main driver of the stellar population gradients 
in both, early, and late-types galaxies. 
Analogously, \citet{Zheng:2017} find that the mean age, and metallicity gradients are 
small however slightly negative. They conclude that their results are consistent 
with the inside-out formation scenario. These authors also study the environmental 
dependence of age and metallicity at the effective radii, finding that high-mass 
galaxies are less affected by the environment.

This paper is the second in a series. In the first paper \citep{Coenda:2019}, we 
explore the effects of environment on the star formation in late-type galaxies, by 
analysing the radial profiles of the specific star formation rate (sSFR). In that 
paper, we consider three different environments: field galaxies, 
galaxies in pairs, and galaxies in groups. Galaxies were selected from the Calar 
Alto Legacy Integral Field Area (CALIFA). 

This article aims to study the effects of the environment over a longer time scale,
analysing the stellar metallicity gradients. For this analysis, we use the subsamples 
of galaxies in groups and in the field, selected by \citet{Coenda:2019}.
This paper is organized as follows: in Sect. \ref{data} we describe in detail
our data, namely, the CALIFA data, the environment classification, and 
the metallicity profiles. In Sect. \ref{results} we present our analysis of the
metallicity profiles of late-type galaxies in the field and in groups.
Finally, we discuss our results in Sect. \ref{Conclusion}.

\section{The Sample}\label{data} 

\subsection{CALIFA}

One of the the most important integral field surveys of the last decade considering 
the sample size and the compromise between Field-of-View, spectral coverage, and 
spatial resolution is the CALIFA Survey. Over five years, more than 900 galaxies were 
observed. The reduced data, ready for scientific exploitation, were made public in 
three successive public releases 
(DR1, \citealt{Husemann:2013}; DR2, \citealt{Garcia-Benito:2015}; DR3,
\citealt{Sanchez:2016}). The instrument used was the 
Potsdam Multi-Aperture Spectrograph \citep[PMAS,][]{Roth:2005}  in the PPaK integral 
field mode \citep{Kelz:2006}, mounted on the $3.5{\rm m}$ telescope of Calar Alto 
Observatory. CALIFA observations were performed in two different configurations. On 
the one hand, using the low resolution V500 grating covering the wavelength range 
$3745 - 7500\AA$, with a spectral resolution of $6.0 \AA$ (Full Width Half Maximum, 
FWHM). The other configuration used the V1200 grating with a medium spectral 
resolution of $2.3 \AA$ (FWHM), in the wavelength range $3650 - 4840\AA$. The total 
number of galaxies observed in this second configuration were 484. From the 
combination of these two configurations, a third data cube called COMBO is obtained 
with a spectral resolution of $6.0 \AA$, and a wavelength range between $3700 - 
7500\AA$. DR3 made available COMBO data cubes for 446 galaxies.

To construct the CALIFA mother sample, 997 galaxies were selected from the Sloan Digital Sky Survey Data Release 7 photometric galaxy catalogue \citep[SDSS DR7,][]{dr7}. With an $r$-band isofotal angular diameter between $45'' - 80''$, covering the redshift range $0.005 < z < 0.03$, the CALIFA mother sample covers the color-magnitude diagram and probes a wide range of stellar masses, ionization conditions, and morphological types. The sample morphological classification was performed through visual inspection of the SDSS $r$-band images by five members of the collaboration. Complete information on the sample characterisation can be found in \citet{Walcher:2014}. More details on the reduction, observation, and data can be found in \citet{Husemann:2013}, \citet{Garcia-Benito:2015}, and
\citet{Sanchez:2016}.

\citet{deAmorim:2017} analyzed all DR3 galaxies with COMBO datacubes using the STARLIGHT\footnote{http://starlight.ufsc.br} spectral synthesis code \citep{CidFernandes:2005}. In combination with the organization code PyCASSO\footnote{Python CALIFA Starlight Synthesis organizer, http://pycasso.ufsc.br, mirror at http://pycasso.iaa.es} \citep{Cid:2013}, they built a catalogue of the stellar populations properties of 445 CALIFA galaxies, providing integrated properties and maps of the stellar mass surface density, mean stellar ages and metallicities, stellar dust attenuation, star formation rates, and kinematics. To perform the analysis of star populations, \citet{deAmorim:2017} used two sets of single stellar population (SSP) bases called GMe and CBe, similar to the GM and CB bases presented in \citet{CidFernandes:2014}, but extended in terms of metallicity coverage. As in \citet{Coenda:2019}, we use in this paper the maps generated with the GMe base which is constructed with a combination of 235 SSP spectra by \citet{Vazdekis:2010}, for populations over 63 Myr, and the models by \citet{GonzalezDelgado:2005}, for younger ages. The initial mass function used is that of \citet{Salpeter:1995}, and the metallicity covers the seven bins $(Z/Z_{\odot}) =-2.3, -1.7, -1.3, -0.7, -0.4, 0, +0.22$ \citep{Vazdekis:2010} for SSP over $63 {\rm Myr}$, and only the four largest metallicities for younger SSP. The evolutionary tracks are those of \citet{Girardi:1993}, except for younger ages (1 and 3 ${\rm Myr}$) for which
Geneva tracks \citep{Schaller:1992} are used instead.

In the present work we study the effects of environment on the stellar metallicity radial profiles of a sample of late-type galaxies with stellar masses in the range $9\leq \log(M_{\star}/M_{\odot}) \leq 12$, where $M_{\star}$ is the total stellar mass obtained from the integrated spectra as indicated in \citet{deAmorim:2017}. The sample analyzed in our work consists in a subsample
drawn from the DR3 galaxies that have the COMBO data cube and stellar metallicity maps determined. 

\subsection{Environments}

The sample of late-type galaxies used in this paper are a subset of the 
samples used in \citet{Coenda:2019}, namely, those that have available metallicity
maps, and are either in groups or in the field. We have excluded for the present
paper galaxies in pairs, given the small number of them that have stellar 
metallicity maps, which made it meaningless to split them into as many different 
mass bins as we do with group and field galaxies in the next section.
Thus, after excluding those galaxies in pairs, we analyse the metallicity
profiles of late-type galaxies in groups and in the field. 

The three environments explored in \citet{Coenda:2019} are defined in terms
of a tracer sample of galaxies from the SDSS-DR12 \citep{dr12}, with measured 
redshifts, and restricted to  $r-$band Petrosian magnitudes $r\le 17.77$.
Since the spectroscopic sample of the SDSS is incomplete in redshift
for galaxies brighter than $r=14.5$, in \citet{Coenda:2019} the tracer sample
is improved with the inclusion of all galaxies in 
the DR12 photometric database that have no redshift measured by SDSS, but have 
available redshift in 
the NED\footnote{The NASA/IPAC Extragalactic Database (NED) is operated by the Jet 
Propulsion Laboratory, California Institute of Technology, under contract with the 
National Aeronautics and Space Administration, https://ned.ipac.caltech.edu/} 
database. As CALIFA only observed nearby galaxies ($z<0.03$), 
this addition from the NED database improves the level of completeness of the tracer
sample, with the consequent improvement of environment characterisation.
We describe briefly how groups and field galaxies are defined in 
\citet{Coenda:2019}.

\subsubsection{Galaxies in groups} 

Our sample of galaxies in groups includes all CALIFA galaxies that
are members of one of the groups of galaxies identified over the tracer sample. 
Groups of galaxies were identified following \citet{Merchan&Zandivarez:2005}. 
For details of group identification we refer the reader to that paper.
In brief, the group sample was constructed by means of the algorithm developed by 
\citet{H&G:1982} that groups galaxies into systems using a redshift-dependent 
linking length. This linking length is tuned to retrieve regions with a numerical 
overdensity of galaxies of 200.  A lower limit in membership is imposed, excluding 
groups with less than four galaxy members. Line-of-sight velocity dispersions are 
computed using the Gapper estimator in the case of groups with less than 15 
members, while for richer groups the bi-weight estimator is used instead 
\citep{Girardi:1993,Girardi:2000}. Group virial mass is estimated through 
the velocity dispension and the projected virial radius.
The resulting sample comprises 17,021 groups with at least four 
members in the redshift range $0<z<0.3$, their virial masses range from 
$\sim 1\times 10^{10} M_{\odot} $ to $ \sim 1\times 
10^{16} M_{\odot} $ with a median of $\sim 8.5\times 10^{13} M_{\odot} $.
A total of 204 CALIFA galaxies are found to be in these groups, 
among them, 112 are late-types.

\subsubsection{Field galaxies}

We consider as field galaxies those CALIFA galaxies that are not included in the
group galaxies defined above, and also that are not likely to be part of a pair
of galaxies. We explain briefly now, how pairs were defined in \citet{Coenda:2019}. Firstly, we search among CALIFA galaxies not included in groups,
those that are candidate to be in a pair. These are galaxies that have a tracer
companion inside a line-of-sight cilinder centered in the CALIFA galaxy and 
that extends out to a projected radius of $100~{\rm kpc}$, and stretches $\pm 
1000~\kms$ in radial velocity \citep{Alpaslan:2015}. This results in a candidate list 
including 127 galaxies, out of which 104 are late-types.

Secondly, a genuine pair is considered as having the line-of-sight 
relative velocity of the two galaxies smaller than the escape velocity of a suitable
dark matter halo (see details in \citealt{Coenda:2019}) at a distance equal to the 
projected separation of the galaxies (see also \citealt{Sales:2007}).
A total of 77 CALIFA galaxies meet this criterion, out of which 62 are late type.
We note that \citet{Barrera-Ballesteros2015b}, use a less conservative set of parameters than in the present work for defining pairs from the CALIFA sample.

The remaining CALIFA galaxies that were not classified as being part of either, 
a group or a pair, are considered as field galaxies. These amount to 226 galaxies, 
including 185 late-types. Some of these galaxies may not be isolated, but 
in actual in pairs or groups that our procedure has not been able to detect.
Thus, the differences we find in the analyses below between field galaxies and 
group galaxies could actually be more significant. 
It is, however, unlikely that this possible contamination from
galaxies in pairs or groups could change our conclusions.

   \begin{figure*}
   \centering
   \includegraphics[width=\hsize]{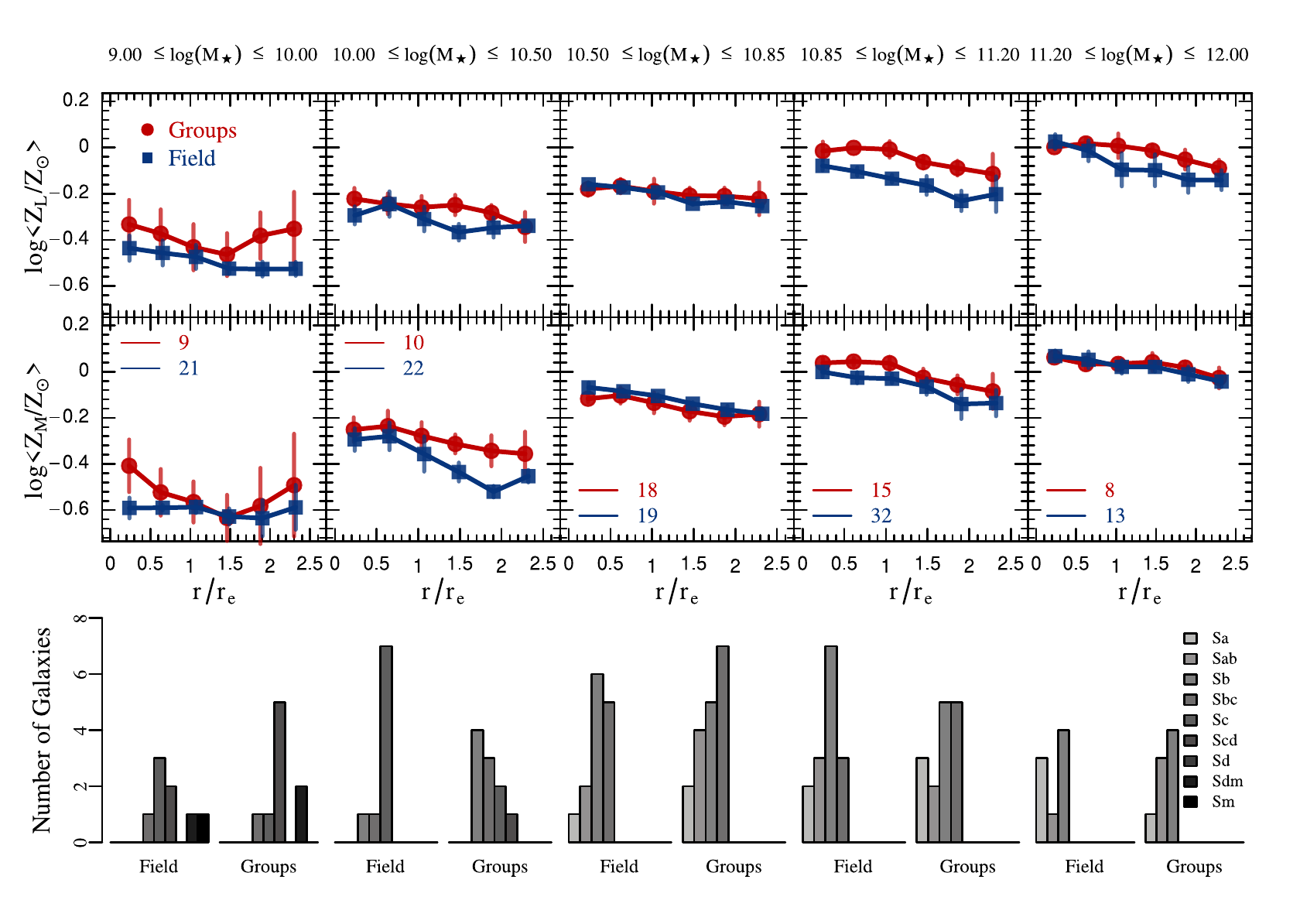}
      \caption{The stacked profile of the metallicity weighted by light (\emph{top panel}) and mass (\emph{central panel}), for late-type galaxies, scaled to the $r$-band half-light effective radius, as a function of the mass and the environment. Galaxies in groups are shown as red dots and lines, and field galaxies as blue dots and lines. Red symbols represent the median in each radial size bin for galaxies in groups. Vertical error bars were computed by using the bootstrap re-sampling technique. The blue symbols show the mean value and its dispersion of the field galaxies for 50 randoms runs. 
      \emph{Bottom panel} show the distribution of the Hubble type as a function of the environment and stellar mass. For field galaxies we show the mean distribution for 50 randoms runs performed.}
         \label{fig:metal}
   \end{figure*}

\subsection{Radial profiles of luminosity and metallicity}\label{profiles}
 
Following \citet{Coenda:2019},  for the radial profiles determination,  firstly we fit ellipses to the luminosity surface density maps ($\mathcal{L}_{5635\AA}$) provided by \citet{deAmorim:2017}. These maps were constructed by directly measuring the average flux of the spectra in the spectral window of $(5635\pm45)\AA$. Using the task ellipse \citep{Jed:1987} within IRAF\footnote{http://iraf.noao.edu/} with 1 spaxel step ($1''$) we obtain the ellipses that we will use later to obtain the metallicity profiles.
\citet{deAmorim:2017} made available two sets of metallicity maps  (measured in solar units), one weighted by luminosity, and the other weighted by mass.  We use the ellipses obtained as indicated above as input for a new run of the task ellipse over the metallicity maps. In this way we obtain the metallicity profiles, weighted by mass ($\log(Z_{\rm M}/Z_{\odot}))$ and luminosity ($\log(Z_{\rm L}/Z_{\odot}))$, that we use to carry out the analyses presented in this paper. We also use the stellar mass surface density, $\Sigma_{\star}$. in units of ${M_{\odot}}$ pc$
^{-2}$, calculated from the masses derived with STARLIGHT and provided by \citet{deAmorim:2017}. This quantity measures the mass currently trapped in stars, as it was corrected for the mass that returned to the interstellar medium during stellar evolution. Our final sample of late-type galaxies with radial metallicity maps comprises 60 galaxies in groups and 107 galaxies in the field. Median values of stellar mass are $\log(M_{\star}/M_{\odot})=10.72$, and $10.70$, for groups, and field galaxies, respectively. As a sanity check of the ellipse step considered, we have redone our analysis with a $2''$ step instead and found that the median of the radial profiles are not altered substantially, thus all conclusions derived here are maintained.

 \section{Results}\label{results}
 
 \begin{figure*}
  \centering
  \begin{minipage}[c][\width]{0.33\textwidth} 
    \includegraphics[width=\linewidth]{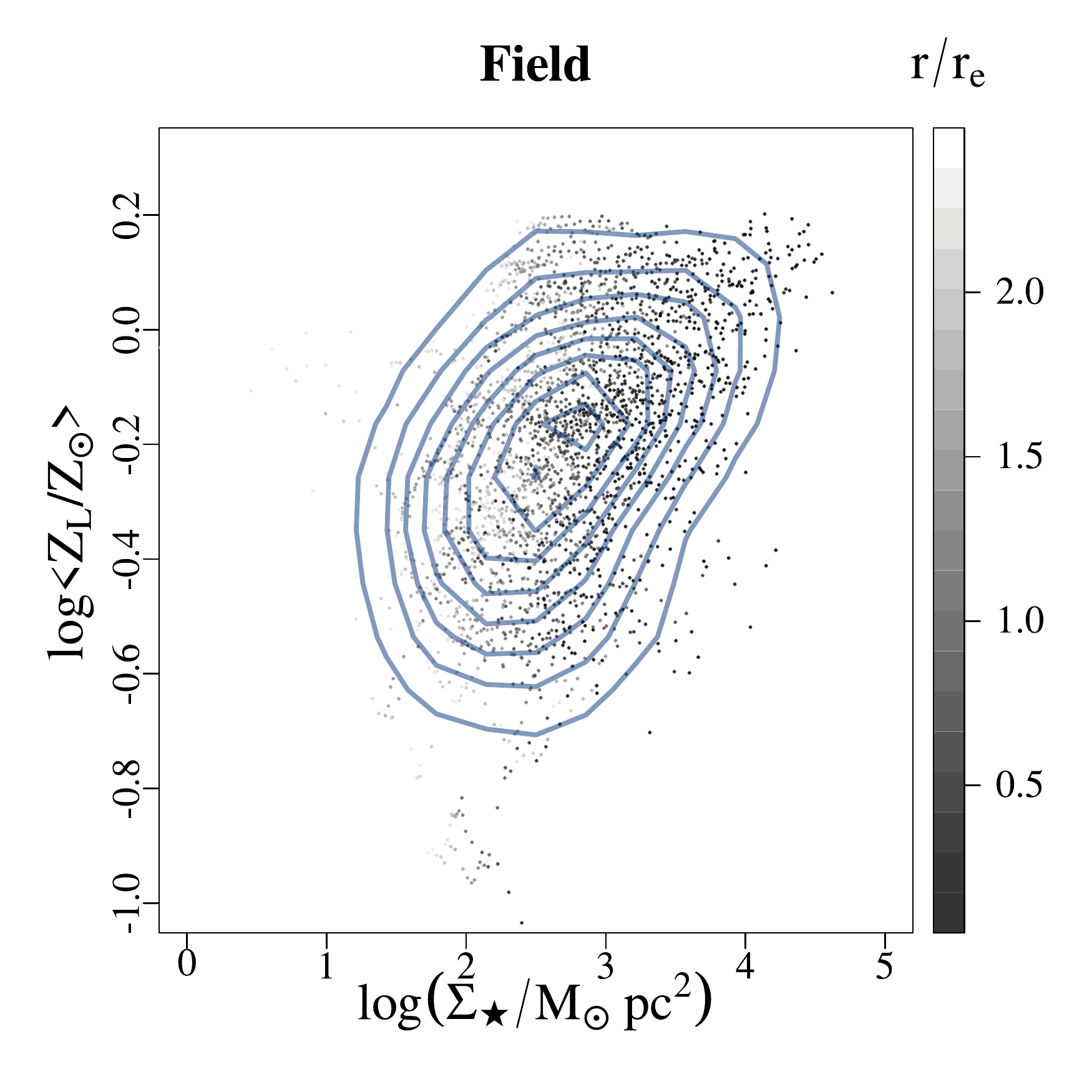}
  \end{minipage}
  \hfill
  \begin{minipage}[c][\width]{0.33\textwidth}  
    \includegraphics[width=\linewidth]{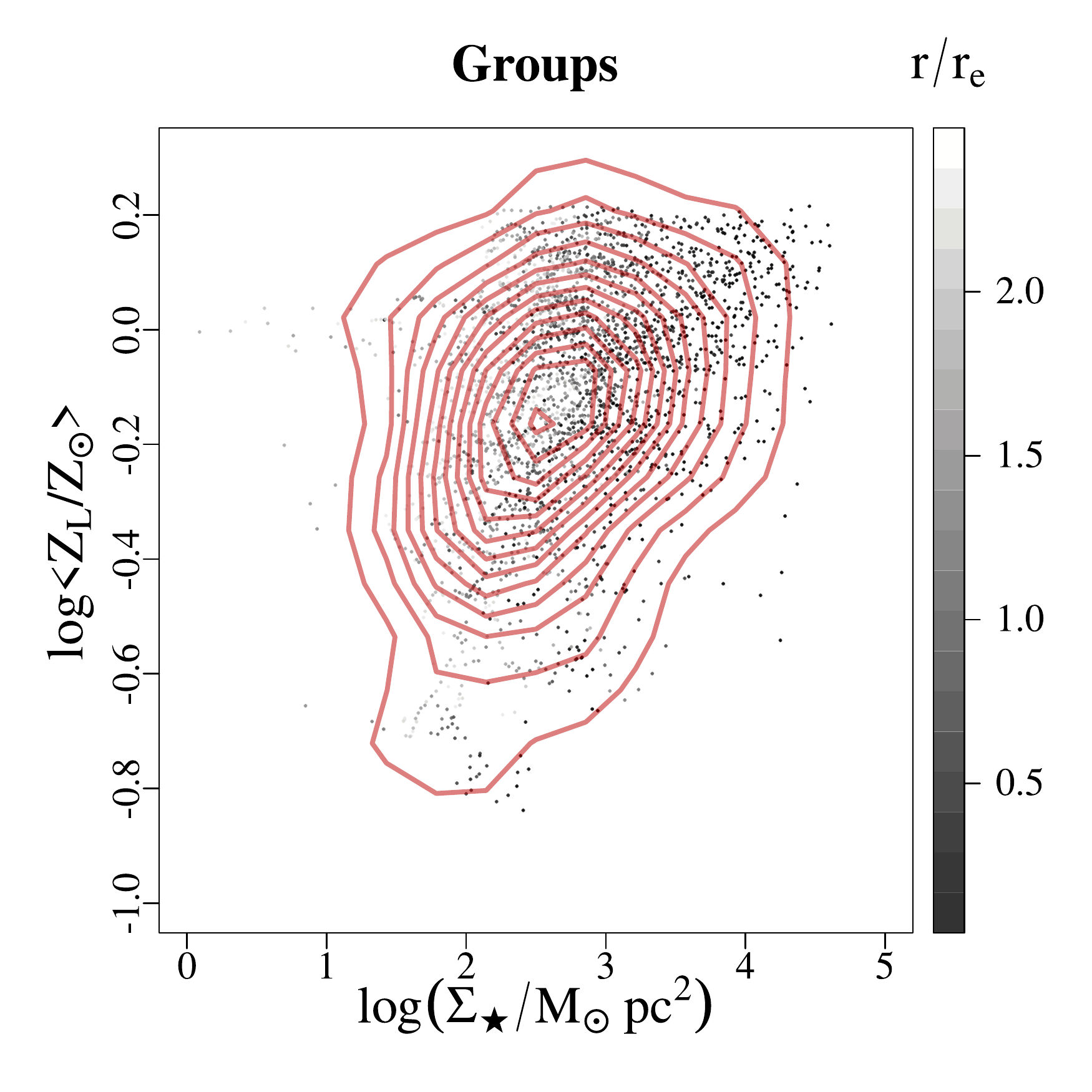}
  \end{minipage}
   \hfill
   \begin{minipage}[c][\width]{0.33\textwidth} 
    \includegraphics[width=\linewidth]{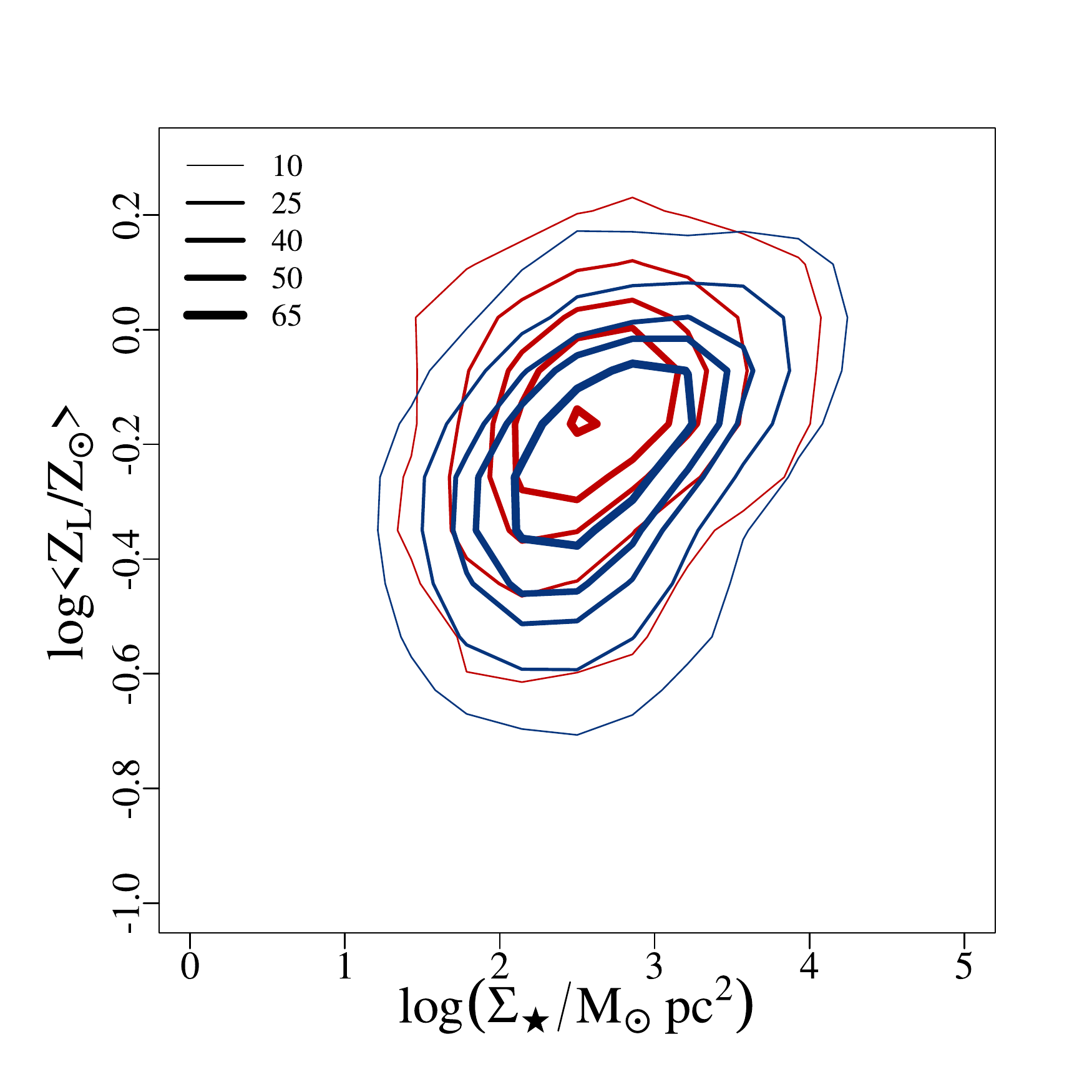}
  \end{minipage}
  \vfill
   \begin{minipage}[c]{\textwidth}
    \includegraphics[width=\linewidth]{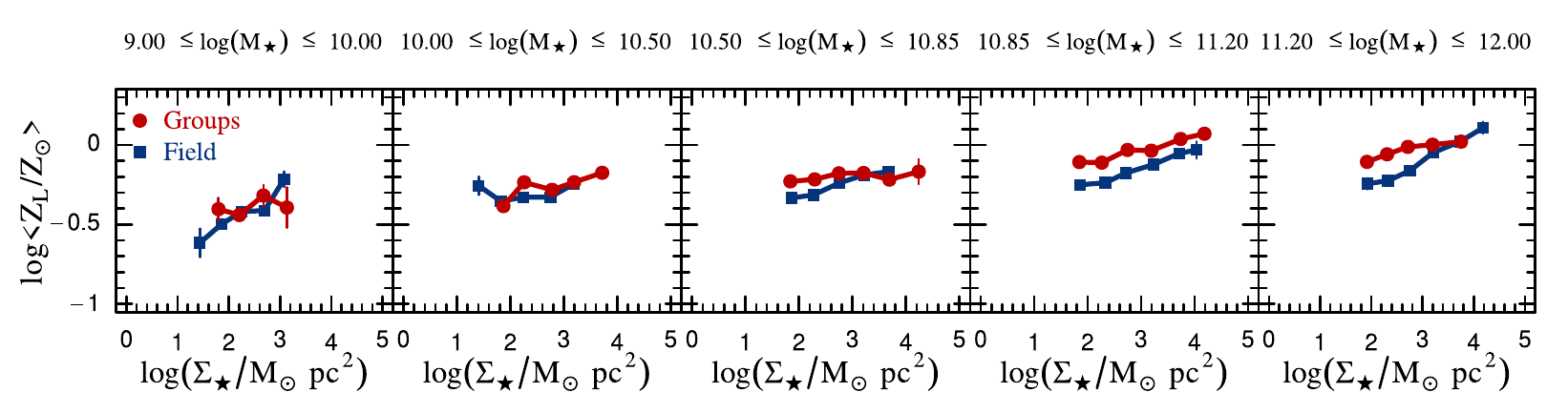}
  \end{minipage}
\caption{Top panels: the light-weighted metallicity as a function of the
stellar mass surface density $\Sigma_{\star}$. Dots are color-coded 
in tones of gray according to their distance $r/r_e$. Left Panel corresponds to galaxies in the 
field, central panel shows galaxies in groups, and right panel shows the 
iso-contour levels for galaxies in groups (red lines), and galaxies in the 
field (blue lines). The bottom panels show the median value of 
$\log(Z_{L}/Z_{\odot})$ as a function of $\Sigma_{\star}$, for the five 
stellar mass bins considered. Vertical error bars were computed by using the bootstrap re-sampling technique.} 
\label{fig:ZM_sup}
\end{figure*}

We study the effects of external and internal mechanisms on the radial distribution of the metallicity for late-type galaxies. To explore the external effects, we compare late-type
galaxies in two discrete environments: field galaxies, and galaxies in groups. To analyse the internal processes, for which mass is the main factor, we split our samples of galaxies into five bins of stellar 
mass:  $\log(M_{\star}/M_{\odot})=9.00-10.00$, $10.00-10.50$, $10.50-10.85$, $10.85-11.20$, and $11.20-12.00$.

Within each stellar mass bin, the samples of field and group
galaxies have, in general, different mass distributions. To avoid mass-related biases in our comparison, we construct subsamples of 
field galaxies (the largest of our samples) randomly selected to have a similar mass distribution
within each mass bin to that of the group sample.
This procedure was performed 50 times for the field galaxies. 

Fig. \ref{fig:metal} compares radial profiles of light- and mass-weighted stellar
metallicity, of galaxies in the field and in groups, split into the five bins of 
stellar mass mentioned above. The spatial scale is the radius in units of 
the the $r$-band half-light effective radius. This effective radius is computed
following \citet{Graham:2005}, involving the SDSS $r-$band 
radius that encloses half the Petrosian flux, and the concentration parameter in 
the same band. We consider the range $0-2.5 r_{\rm e}$ to stack the radial 
profiles and we calculate the median value of $r/r_{\rm e}$ within each 
interval of size. 
 The subscript $L$ and $M$ correspond to the metallicity weighing by light and
 mass, respectively. For galaxies in groups, Fig. \ref{fig:metal} shows the 
 median of $\log(Z_{\rm L(M)}/Z_{\odot})$ as a function of $r/r_e$. Vertical 
 error-bars were computed using the bootstrap re-sampling technique. For galaxies 
 in the field, Fig. \ref{fig:metal} shows the mean value of the medians of 
 $\log(Z_{\rm L(M)}/Z_{\odot})$ as a function of $r/r_e$, averaged over the 50
 random realisations. Error-bars in this case are the dispersion around the mean
 value.
Analogously to \citet{GonzalezDelgado:2016} and \citet{Coenda:2019}, we have considered only mass bins containing more than five galaxies. We quote in all cases the actual number of galaxies contributing to each profile.

In general, we observe in Fig. \ref{fig:metal} that $\log(Z_{\rm L(M)}/Z_{\odot})$ increases with the stellar mass, and profiles have negative gradients, as has been reported by other authors (e.g.  \citealt{SanchezBlazquez:2014}, \citealt{GonzalezDelgado:2016}, \citealt{Goddard:2017b}, \citealt{Zheng:2017}, \citealt{Lian:2018}). 
We find that group galaxies are 
systematically more metallic than their field counterparts. This is found 
regardless whether the metallicity is mass-, or luminosity-weighted. For the third 
stellar mass bin, and forboth: the profiles weighted by mass and by luminosity, we observe that galaxies in groups and in the field have similar stellar metallicities. We observe the same trend for the profiles weighted by mass, in the fifth stellar mass bin. With this exceptions, the differences between galaxies in groups and in the field are more noticeable in the luminosity weighted metallicity. Therefore, in what follows, we will centre our analyses in the light-weighted profiles. 
For mass bins second, fourth and fifth, there is a tendency of the median profiles of the metallicity in groups
to have a more flattened metallicity gradient with a negative slope in the outer parts,
whereas for field galaxies a more lineal profiles are observed.  This results in field and group galaxies having roughly similar metallicities at both extremes: the innermost and the outermost regions. The first mass 
bin shows a different behavior.  Although in this case we also observe that group galaxies have higher metallicity than galaxies in the field, the metallicity  profile of the former shows a different shape compared to the other mass bins. In this bin, the metallicity profile of group galaxies presents a convex shape. As a consequence, the differences between the outer and inner radial zones are maximum, and in the middle, the stellar metallicities of both, group, and field galaxies, are similar.
Galaxies in this mass bin have the lowest surface brightness. Although we have checked that all galaxies in the bin contribute to the whole range  of $r/r_e$ probed, caution should be taken regarding the behaviour in the outermost parts because of a possible low S/N effect.

If we assume that field galaxies have evolved virtually in isolation, or at least, that they have suffered much lesser environmental effects than group galaxies, the differences we observe in the metallicity profiles can be related directly to environmental action. 

In the bottom panel of Fig. \ref{fig:metal} we show the distribution of Hubble types of our sample of late-type galaxies, as a function of stellar mass and environment. For field galaxies we show the average distribution of the 50 randoms runs performed.
 We observe the morphology distributions are quite similar in both environments for each stellar mass bins. The only probable exception is the second bin of mass, where we observe a tendency of group galaxies to have earlier morphologies  than field galaxies. This characteristic of the sample may be partially responsible of the observed differences in metallicity between the two environments. It should be noted that this is the bin of mass that presents the larger difference in metallicity. We also note that each stellar mass bin implies a different set of morphologies. As stellar mass increases, a higher fraction of Sa and Sb galaxies is observed.
It is worth noting that the mass-dependent morphological mixing in our
samples implies we are not probing an unique galaxy class across the stellar mass range. On the
contrary, galaxies in each bin constitute a completely independent sample, and the only
common feature bin-to-bin is that galaxies are late-types.

In \citet{Coenda:2019} we explore whether AGN feedback, or the presence 
of a bar, play a role in shaping the sSFR profiles. In contrast to SFR, 
which involves short timescales, in this work we are studying 
the profiles of the stellar metallicities, where different physical 
mechanisms have acted at different temporal and spatial scales 
throughout the galaxy lifetime.
The size of our sample does not allow for a separate analysis on whether 
AGNs, or bars, can play a role in the observed stellar metallicity. We 
observe, however, that the mere presence of an AGN at $z=0$ does not 
necessarily imply that it has played a role in shaping the observed 
galaxy stellar metallicity. 
An AGN can be a transient phenomenon in a galaxy. With AGN duty cycles 
spanning timescales in the range $10^6-10^8$ yr 
\citep{Haehnelt1993,Davis2014,Storchi-Bergmann2019}, we expect that the
AGN effects on stellar metallicity should be due to past time, not 
current, AGN activity. Regarding bars, there is no consensus in the 
literature as to how long do bars last, nor how many bar events an 
average galaxy has during its lifetime \citep{Sellwood1999,Athanassoula2002,Bournaud2002,Elmegreen2004,Combes2004,ReganTeuben2004,Perez2008,James2016}. While several authors have observed a correlation between gas abundance gradients and the presence of bars, 
in particular, a flattening of the gradient 
\citep{VilaCostas1992,MartinRoy1994,Zaritsky1994}, more recent works have 
found no evidence of such a correlation by analyzing both, gas-phase metallicity, and stellar metallicity \citep{Sanchez2012b, Sanchez2014,SanchezBlazquez:2014,Cheung2015,SanchezMenguiano2016}.

In the upper panels of Fig. \ref{fig:ZM_sup} we show the light-weighted
metallicity as a function of the stellar mass surface density
$\log(\Sigma_{\star})$, for galaxies in the field and galaxies in groups.
Field galaxies shown in Fig. \ref{fig:ZM_sup} are a random realization of 107 
galaxies from the field sample, selected to have the same overall 
mass distribution as the group sample. 
Each galaxy, in any of the two samples, contributes several points to 
this figure. The number
of points vary from galaxy to galaxy, as does the number of radial bins
that each profile has. Points are colour-coded in tones of gray according 
to their radial distance in terms of the effective radius, $r/r_{\rm e}$. 
The solid curves show the number-density iso-contour levels. 
The stellar mass surface density 
and metallicity are strongly correlated although we observe significant 
scatter. 
To compare galaxies in the field and groups, we show in the upper right 
panel of Fig. \ref{fig:ZM_sup}, the iso-contour for each environment  
considered. We clearly observe that galaxies in groups are more metallic for a 
fixed value of stellar mass surface density. 

For a better comparison between group and field galaxies, 
we split both samples into the 5 mass bins used in Fig. 
\ref{fig:metal} and  compute the median value of the $\log(Z_{\rm L}/Z_{\odot})$
as a function of the stellar mass surface density, $\log(\Sigma_{\star})$.
This is shown in  the bottom panels in Fig. \ref{fig:ZM_sup}. 
Clearly, metallicity depends on both, mass, and stellar mass surface 
density, i.e., a global, and a local property, respectively.
Mass being a major source of scatter in the top panels of this figure.
We observe no clear distinction between field and group galaxies in the 2 lowest
mass bins. From the third bin onward, we observe that group galaxies tend to 
be more metallic at low to intermediate $\log(\Sigma_{\star})$ values, although it is not clear that the differences are significative in the third stellar mass bin.
Another 
interesting feature in this figure is that field galaxies tend to have a 
stronger dependence of metallicity on  stellar mass surface density. 

A strong correlation between the stellar metallicity and the stellar mass surface
density has been reported by \citet{Rosales-Ortega:2012}, \citet{Sanchez:2013}, 
\citet{GonzalezDelgado:2014}. This correlation can be considered as a local process
acting in galaxies. Previously, \citet{Bell:2000}, analysed spiral galaxies and 
found that the stellar mass surface density of galaxies drives its star formation 
history, and that $M_{\star}$ is a less important parameter. These previous works 
argue that stellar metallicities are mainly governed by the stellar mass surface 
density in disk galaxies, and by the total mass in spheroids. Our results suggest 
that both, stellar mass surface density, and the integrated stellar mass, 
impact on the star formation history of late-types galaxies.  
Moreover, we also find that the environment also plays an important role in modelling the metallicity profiles. 

   \begin{figure}
   \centering
   \includegraphics[width=\hsize]{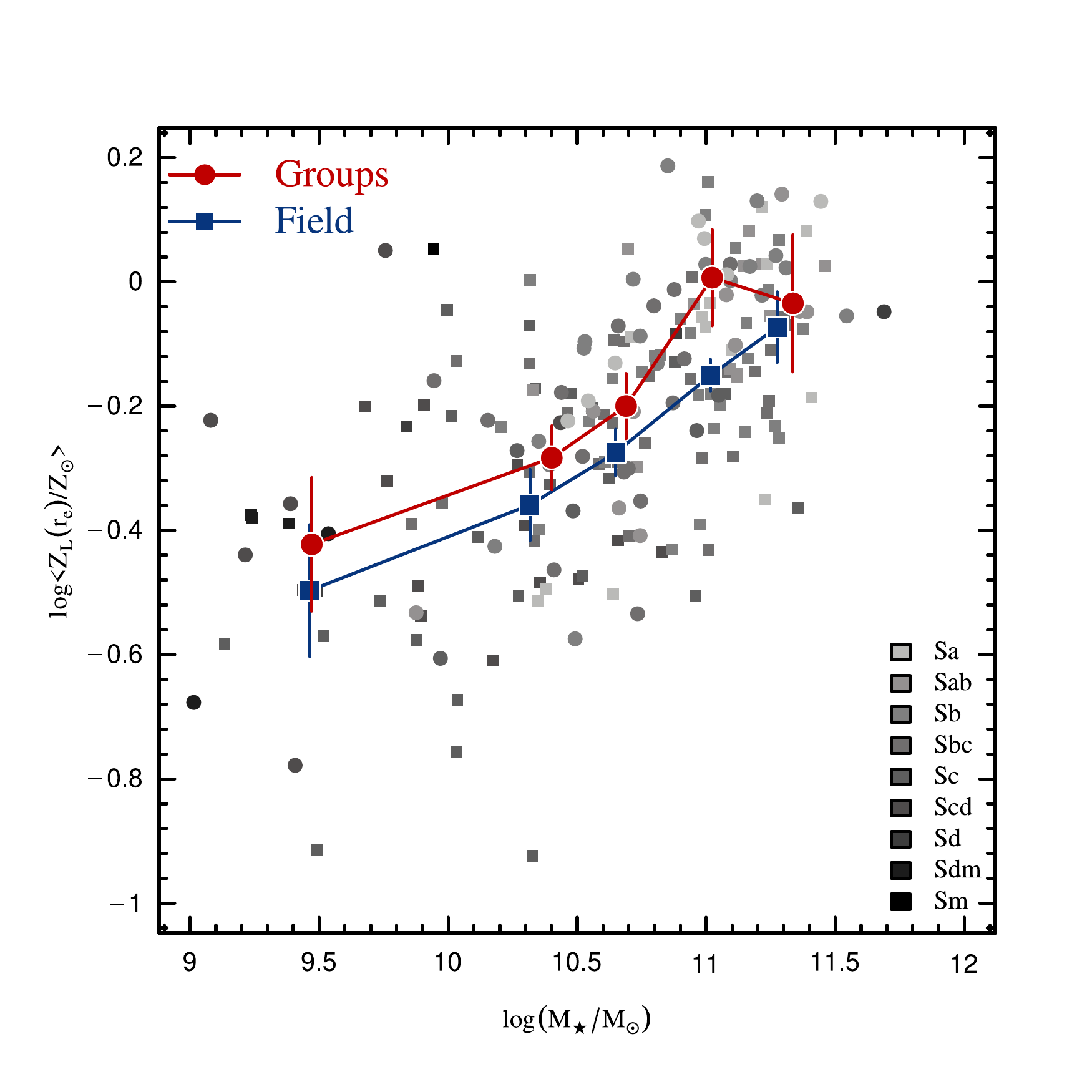}
      \caption{Light-weighted metallicity at $r_{\rm e}$ as a function of the stellar mass. Dots are colour-coded in tones of gray according to their Hubble type. The red lines and dots show the median values of $\log(Z_{L}(r_e)/Z_{\odot})$ for galaxies in groups in red, while galaxies in the field are shown in blue.}
         \label{fig:re}
   \end{figure}
 
A complementary quantity from our profiles is the metallicity value at 
$r_e$. We show in Fig. \ref{fig:re}, the median light-weighted stellar 
metallicity at $r_e$, as a function of the stellar mass of the galaxy. The
metallicity at $r_e$ is correlated with $M_{\star}$ for late-type 
galaxies, and this correlation depends on the environment. 
Low-mass galaxies have lower metallicity than high-mass galaxies 
Again, galaxies in groups show a higher value of metallicty than galaxies in 
the field, for a fixed value of stellar mass. \citet{Zheng:2017} find that low-mass galaxies tend to have lower metallicity in low-density environments while high-mass galaxies are less affected by environment. Our results suggest that the environment affects the metallicities in the whole range of stellar mass, however, future analyses with larger samples are necessary.

\section{Discussion and conclusions}\label{Conclusion}

In this paper we present a comparative analysis of the stellar metallicity profiles
of late-type galaxies in the field and in groups, using publicly available CALIFA data.
We focus on three comparative analyses of the metallicity of late-type galaxies: 
(i) the metallicity profiles in five stellar mass bins, (ii) the relation between metallicity and
the stellar mass surface density, (iii) the metallicity at the effective radius.
Thus, our analyses compare the metallicity of galaxies as a function of scale (i), at a
characteristic scale (iii), and as a function of a local property (ii).
As most galaxy properties, including metallicity, depend on galaxy mass, we take special care
throughout the paper in order to compare, in all cases, subsamples of galaxies in groups and in 
the field that have similar mass distributions, within the mass ranges analysed, however 
broad or thin these ranges might be.
In all cases we find significant differences between group and field late-type galaxies.
Our results contrast  with those of  \citet{Goddard:2017a}, who find that stellar
population gradients have no significant correlation with galaxy environment
regardless the different characterisations of environment they use.
This difference between our results and \citet{Goddard:2017a} could be
due to our choice of splitting galaxies into two discrete environments.

Regarding the comparison of the radial profiles of metallicity, we find that field galaxies
have, in general, metallicity profiles that show a negative gradient in their inner
regions, and a shallower profile at larger radii. 
This contrasts with the metallicity profiles of group galaxies, which tend to be flat in the inner regions, and show a negative gradient in the outer parts.
As bulges of late-type galaxies are denser and have stars older than the disk's, they are 
expected to be more metallic than the outer parts of the galaxy.

A plausible scenario could be one in which SN ejections, throughout the lifetime
of a galaxy, are accreted back into the disk in field galaxies, thus increasing the metallicity 
of stars formed later in the outer parts of the disk. This should be less efficient in groups 
due to environmental effects such as ram-pressure stripping or strangulation.
Group galaxies have higher metallicity than field galaxies at most scales, and
notably at the characteristic radius.
Since groups are dense environments, galaxies in groups should, on average,
have been formed earlier (i.e. downsizing), thus having more time to produce 
metals. On the other hand, mergers are common in groups, and they tend to 
redistribute metals within galaxies. Furthermore,
mergers could add metals to galaxies by the accretion of earlier types 
satellite galaxies. 

Analysing the light-weighted stellar metallicity at $r_e$, we consistently find 
that the correlation depends on the environment.  Again, galaxies in groups show a 
higher value of metallicty than galaxies in the field, at a fixed value of 
stellar mass.These results are consistent with the findings of 
\citet{Zheng:2017}, however, in our case the evidence is not only present for low-mass galaxies, but in nearly the entire mass range, being the only exception the highest mass bin, where the  median values of metallicities are indistinguishable.

Our analysis of the dependence of metallicity on stellar mass surface density shows
that, in general, at fixed local density, group galaxies are more metallic, which
backs up the idea that they formed earlier. Another general trend is that the dependence
of metallicity on surface density is less important in group galaxies, which may
be an indicative of a more effective mix, that can be thought of in terms of more
frequent mergers throughout their lifetimes. 

As seen in previous works, $\Sigma_{\star}$ is a good tracer of the local star population properties where both age and metallicity correlate with $\Sigma_{\star}$
\citep{Rosales-Ortega:2012,Sanchez:2013, GonzalezDelgado:2014}. In particular, in disks $\Sigma_{\star}$ would regulate the mean stellar ages and metallicities, while in spheroids, both in spiral bulges and elliptical galaxies, $\Sigma_{\star}$ would play a minor role. This is due, as mentioned above, to the fact that in spheroids the chemical enrichment occurred faster and at an earlier stage than in disks, in full correspondence with the inside-out scenario \citep{Perez:2013,GonzalezDelgado:2014,GonzalezDelgado2015,SanchezBlazquez:2014,Sanchez2014,Garcia-Benito:2017}. This would imply that stellar metallicity would be governed 
by local processes in disks and by global processes in spheroids 
\citep{Gonzalez-Delgado2016}. Our $\Sigma_{\star}$ analysis shows how, from medium to 
high densities, which we could associate with spheroids, the effect of the environment 
is diluted, while the major differences between groups and field occur as 
$\Sigma_{\star}$ decreases. This effect is evidently more noticeable in the medium to 
high mass bins, bins in which, due to the morphological distribution of the sample, 
there is a more significant presence of spheroids. This could indicate, therefore, that 
the environment plays an important role in the chemical evolution of disks and perhaps 
a minor role in spheroids.

Alongside strong evidences of the universality of the inside-out formation of galaxies, \citet{Garcia-Benito:2017} have shown a complex multivariate dependence of the
mass assembly on stellar mass, stellar mass surface density, and Hubble 
type. In \citet{Coenda:2019}, and in this paper, we have taken special care of these
factors in our analyses, and have shown that environment 
is another factor to reckon with at the time of understanding in detail 
how galaxies form. In this paper in particular, we have presented
evidence of a clear difference on the metallicity of group and
field galaxies, as a function of mass, spatial scale, and local stellar mass density.
From an earlier start, a mayor number of mergers experienced during their lifetimes, and the action of other environmental mechanisms, it is clear that late-type galaxies in groups 
have followed a different evolutionary path, compared to their field counterparts.

\begin{acknowledgements}
This paper is based on data obtained by the CALIFA survey
(http://califa.caha.es) which is based on observations collected at the Centro Astron\'omico Hispano Alem\'an (CAHA) at Calar Alto, operated jointly by the Max-Planck-Institut f\"ur Astronomie and the Instituto de Astrof\'{i}sica de Andaluc\'{i}a (CSIC). 
This research has made use of the NASA/IPAC Extragalactic Database (NED), which is operated by the Jet Propulsion Laboratory, California Institute of Technology, under contract with the National Aeronautics and Space Administration.
This paper has been partially supported with grants from Consejo Nacional de 
Investigaciones Cient\'ificas y T\'ecnicas (PIP 11220170100548CO) Argentina, Fondo para la Investigación Científica y Tecnológica (FonCyT, PICT-2017-3301), and Secretar\'ia de Ciencia y Tecnolog\'ia, Universidad Nacional de C\'ordoba, Argentina.
\end{acknowledgements}



\end{document}